\documentclass[reprint, aps, superscriptaddress, prb]{revtex4-2}
\usepackage{graphicx}
\usepackage{float}
\usepackage{amsmath}
\usepackage{amsfonts}
\usepackage{svg}
\usepackage{physics}
\usepackage{commath}
\usepackage{multirow}
\usepackage[utf8]{inputenc}
\usepackage[margin=1in]{geometry}
\usepackage{comment}
\usepackage{dcolumn}
\usepackage{bm}
\usepackage{textgreek}
\usepackage{xr-hyper}
\usepackage{hyperref}
\usepackage[capitalize]{cleveref}
\makeatletter

\begin{document}

\preprint{APS/123-QED}

\title{Silicon T centre hyperfine structure and memory protection schemes}

\author{Nicholas Brunelle}
\thanks{These authors contributed equally to this work.}
\affiliation{Department of Physics, Simon Fraser University, Burnaby, British Columbia, Canada}
\affiliation{Photonic Inc., Coquitlam, British Columbia, Canada}
\author{Joshua Kanaganayagam}
\thanks{These authors contributed equally to this work.}
\affiliation{Department of Physics, Simon Fraser University, Burnaby, British Columbia, Canada}
\affiliation{Photonic Inc., Coquitlam, British Columbia, Canada}
\author{Mehdi Keshavarz}
\affiliation{Department of Physics, Simon Fraser University, Burnaby, British Columbia, Canada}
\affiliation{Photonic Inc., Coquitlam, British Columbia, Canada}
\author{Chloe Clear}
\affiliation{Department of Physics, Simon Fraser University, Burnaby, British Columbia, Canada}
\affiliation{Photonic Inc., Coquitlam, British Columbia, Canada}
\author{Oney Soykal}
\affiliation{Photonic Inc., Coquitlam, British Columbia, Canada}
\author{Myles Ruether}
\affiliation{Department of Physics, Simon Fraser University, Burnaby, British Columbia, Canada}
\affiliation{Photonic Inc., Coquitlam, British Columbia, Canada}
\author{Adam DeAbreu}
\affiliation{Department of Physics, Simon Fraser University, Burnaby, British Columbia, Canada}
\affiliation{Photonic Inc., Coquitlam, British Columbia, Canada}
\author{Amirhossein AlizadehKhaledi}
\affiliation{Department of Physics, Simon Fraser University, Burnaby, British Columbia, Canada}
\affiliation{Photonic Inc., Coquitlam, British Columbia, Canada}
\author{Yihuang Xiong}
\affiliation{Thayer School of Engineering, Dartmouth College, Hanover, New Hampshire 03755, USA}
\author{Nikolay V. Abrosimov}
\affiliation{Leibniz-Institut für Kristallzüchtung, Berlin, Germany}
\author{Geoffroy Hautier}
\affiliation{Thayer School of Engineering, Dartmouth College, Hanover, New Hampshire 03755, USA}
\affiliation{Department of Materials Science and NanoEngineering, Rice University, Houston, TX, USA}
\affiliation{Rice Advanced Materials Institute, Rice University, Houston, TX, USA}
\author{Michael Thewalt}
\affiliation{Department of Physics, Simon Fraser University, Burnaby, British Columbia, Canada}
\affiliation{Photonic Inc., Coquitlam, British Columbia, Canada}
\author{Stephanie Simmons}
\affiliation{Department of Physics, Simon Fraser University, Burnaby, British Columbia, Canada}
\affiliation{Photonic Inc., Coquitlam, British Columbia, Canada}
\author{Daniel Higginbottom}
\email{Contact author: dhigginb@sfu.ca}
\affiliation{Department of Physics, Simon Fraser University, Burnaby, British Columbia, Canada}
\affiliation{Photonic Inc., Coquitlam, British Columbia, Canada}

\date{\today}

\begin{abstract} 

Combining the long-coherence of spin qubits and the capability to transmit information and entanglement through photons, spin-photon interfaces (SPIs) are a promising platform for networked quantum computation and long-distance quantum communication. SPIs that possess local `memory' qubits in addition to the optically coupled `communication' qubit can improve remote entanglement fidelities through brokered entanglement schemes and entanglement purification. In these schemes, it is critical to protect the memory qubit from decoherence during entanglement operations on the communications qubit. Silicon, a platform with mature microelectronic and nanophotonic fabrication, is host to the T centre, an SPI with emission in the telecommunications O-band that directly integrates with silicon nanophotonics. Cavity-coupled T centres are a platform for brokered entanglement distribution in silicon photonic circuits and over long-distance optical fibre links. The T centre's electron and nuclear spin qubits are an intrinsic register of communication and memory qubits respectively, with anisotropic hyperfine coupling. In this work we determine the T centre's hydrogen hyperfine coupling tensor. We also introduce schemes to protect against dephasing or eliminate relaxation of the T centre's hydrogen memory qubit during optical excitation. These results address a key challenge for practical T centre quantum networks.

\end{abstract}

\maketitle

\section{Introduction}

Quantum networks are the basis of securely encrypted communication through quantum key distribution (QKD) and distributed quantum computation over interconnected modular quantum computers \cite{Whener2018quantum, Benjamin2009prospects, Monroe2014, Li2024High-RateProcessors}. Networked quantum computing can overcome the constraints facing monolithic quantum devices and enables quantum internet applications including blind quantum computing \cite{Broadbent2009universal,Drmota2024VerifiablePhotons}. Quantum networks require coherent quantum links between network nodes with long-lived quantum memory \cite{Childress2005Fault-tolerantEmitters}. Spin-photon interfaces (SPIs) are network nodes combining the long coherence of spin qubits and spin-dependent single-photon emission \cite{Pompili2021a,Stas2022robust,Knaut2024EntanglementNetwork} that can be entangled even over lossy optical links \cite{Barrett2005, Cabrillo1999, Hong1987MeasurementInterference}. 

A spin-photon network is bolstered by the addition of a spin register at each SPI \cite{Childress2005Fault-tolerantEmitters}. With more than one spin qubit at a given node, their roles can be divided into optically-coupled `communication' and long-lived `memory' qubits \cite{Benjamin2006BrokeredComputation}. In such a protocol, the communication qubits implement probabilistic entanglement generation and only interact with the memory qubits to transfer entanglement. Common remote entanglement schemes are intrinsically probabilistic, and losses from the path and detectors further reduce the success probability. Because each entanglement attempt is low probability, the memory qubit must remain coherent over many attempts (as many as $\approx10^4$ in \cite{Humphreys2018}). Brokered entanglement has been demonstrated with trapped ions \cite{Drmota2023robustquantum} and NV centres \cite{Pompili2021a}, and has enabled entanglement over 230~m \cite{Krutyanskiy2023entanglement}, QKD certified by Bell's theorem \cite{Nadlinger2022experimental}, and blind quantum computing demonstrations \cite{Drmota2024VerifiablePhotons}. Quantum memory registers can also be used to generate higher-dimensional optical resource states for one-way quantum computation and communication \cite{Michaels2021, Pieplow2023DeterministicDiamond}.

Solid state emitters such as quantum dots \cite{Kim2011, Press2008}, impurities (Er) \cite{Yin2013, Gritsch2023Purcell}, and colour centres in diamond \cite{Stolpe2023mapping, Knaut2024EntanglementNetwork, Bradley2019AMinute, Stas2022robust}, Si \cite{Bergeron2020, Higginbottom2023memory}, SiC \cite{Falk2015OpticalCarbide, Klimov2015QuantumEnsemble}, or 2D materials (HBN) \cite{Stern2022Room-temperatureNitride} can host optically-coupled communication qubits (usually an electron or a hole) and hyperfine-coupled nuclear spins that may not be optically resolved, but make long-lived memory qubits. Together these can form a central-spin quantum processor \cite{Abobeih2022fault}. Nuclear spins may be distributed randomly throughout the SPI's crystal host, in which case their coupling to the SPI must be characterized individually \cite{Stolpe2023mapping}. Uncharacterized host spins are a source of decoherence. Alternatively, central-spin systems can utilize local nuclear spins that are inherent to the emitter, with a predictable hyperfine coupling. Multi-atom centres may host several nuclear spins, forming a deterministic and intrinsic nuclear spin register \cite{Lockyer2021TargetingCorrection}. 

Although communication-memory qubit coupling is required for local operations, it is also (in general) a source of decoherence during entanglement attempts \cite{Pompili2021a}. Unlike trapped ion network nodes, the memory qubits in solid state SPIs cannot simply be shuttled away from the communication qubits \cite{Drmota2024VerifiablePhotons} to disable the interaction. NV centre network nodes have utilized spin-0 states \cite{Pompili2021a}, weakly-coupled remote spins \cite{Bradley2022RobustQuantum}, and decoherence-free subspace encodings \cite{Reiserer2016Robust} to minimize optically induced decoherence and explored the use of spectator qubits to mitigate dephasing \cite{Loenen2025QuantumNetwork}.

In this work, we consider memory qubits within the silicon T centre, a multi-atom SPI with an inherent spin register. The T centre emits in the telecommunications O-band at 1326~nm \cite{Bergeron2020}. Measurements of this transition have found homogeneous linewidths as narrow as 0.69(1)~MHz \cite{DeAbreu2023Waveguide-integratedCentres}. Being native to silicon, the T centre can be integrated in silicon photonic nanostructures such as waveguides \cite{DeAbreu2023Waveguide-integratedCentres,Lee2023highefficiency} and cavities \cite{Islam2023cavityenhanced,Johnston2024Cavity-coupledSilicon,Dobinson2025ElectricallyTriggered,Bowness2025Laser-inducedCentres, Komza2025MultiplexedArray}. The T centre has long spin coherence times, $>2$~ms and $>1$~s for the electron and nuclear spins respectively \cite{Bergeron2020}. Depending on isotopic composition, the T centre can host up to three spin-1/2 nuclei: two $^{13}$C nuclei plus the hydrogen nucleus. Coupling to nearby $^{29}$Si nuclear spins has also been reported \cite{Song2025LongLived} with hydrogen nucleus effective hyperfine couplings consistent with our presented results for the z0 or z2 orientations as defined by Clear et al. \cite{Clear_2024_OpticalTransitionParameters} for a small ($<4.5^\circ$) misalignment from the $[110]$ axis. The electron has an anisotropic hyperfine coupling to the carbon and hydrogen nuclei, which provides a degree of tunability through the choice of magnetic field direction. The hyperfine coupling tensors have not previously been reported.

In this paper, we characterize the most common, naturally-occurring isotopic variant, with two spin-0 $^{12}$C nuclei and the hydrogen nucleus as the sole nuclear spin. We measure the anisotropic hyperfine interaction between the ground state electron and nuclear spins and unambiguously determine the coupling tensor. We model the magnetic field dependence of memory qubit decoherence under optical excitation. From this, we determine operational conditions preserving the nuclear spin qubit state through many remote entanglement attempts.

\begin{figure}
    \centering
    \includegraphics[]{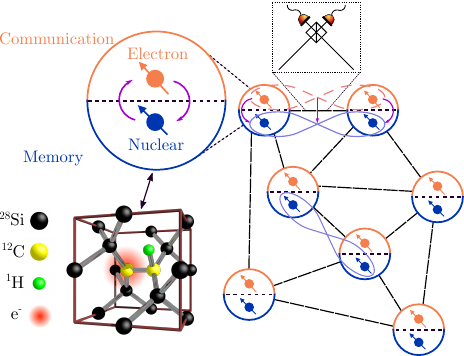}
    \caption{T centre crystal structure and broker-client model. The electron and nuclear spins of the T centre can be assigned communication and memory roles respectively. Each node in the network corresponds to an addressable T centre, and each node can be as close as neighbouring photonic devices or as far as kilometres through telecom fibre connections. Remote entanglement of communication qubits is performed by interfering emission from two connected nodes. This entanglement is then transferred to the memory qubits through the hyperfine coupling (purple arrows) between the electron and nuclear spins.}
    \label{fig:Broker Client}
\end{figure}

\section{T Centre Hydrogen Hyperfine}

The T centre comprises two bonded carbon atoms sharing a silicon lattice site, with a hydrogen atom bonded to one carbon and an unpaired electron in the dangling bond of the trigonal carbon \cite{Safonov1996b} (see \cref{fig:Broker Client}). It has 24 unique orientations that divide into 12 inversion symmetric pairs under applied magnetic fields \cite{KAPLYANSKII1967, Safonov1995, Safonov1996b}. The ground state, T$_0$, comprises a spin-1/2 electron and a spin-1/2 hydrogen nucleus. The T centre can bind an exciton, which recombines to emit a $935$~meV (1326~nm) photon. In the lowest energy bound exciton state, TX$_0$, the electrons form a singlet and the weakly-bound hole is delocalized compared to the T$_0$ electron wavefunction \cite{Dhaliah2022, Clear2024optical, Bergeron2020}. For a given T centre, an external magnetic field will split the zero-phonon line (ZPL) transition from TX$_0$ to T$_0$ into four optically resolvable transitions.

In the ground state, the electron and nuclear spin are hyperfine coupled. The hyperfine splitting at zero magnetic field was measured to be 3.85(1)~MHz by optical hole burning under the assumption that ground state has degenerate triplet states and a singlet state at zero field \cite{DeAbreu2023Waveguide-integratedCentres}. We demonstrate in this work that the triplet states are not degenerate at zero field. Displacement of the electron from the hydrogen nucleus, as depicted in \cref{fig:Broker Client}, yields an anisotropic hyperfine coupling with Hamiltonian
\begin{equation} \label{eq:hyperfine ham}
    H_\text{HF}=\hat{\textbf{S}}\textbf{A}\hat{\textbf{I}}=\sum_{i,j = x,y,z}A_{ij}S_i\otimes I_j\, ,
\end{equation}
where $\hat{\textbf{S}}$ is the electron spin, $\hat{\textbf{I}}$ is the nuclear spin, and $\textbf{A}$ is the hyperfine coupling tensor. The ground state Hamiltonian consists of the electron Zeeman ($H_\mathrm{EZ}$), nuclear Zeeman ($H_\mathrm{NZ}$), and hyperfine ($H_\mathrm{HF}$) components
\begin{equation} \label{eq:Tot Ham}
    H=H_\mathrm{EZ}+H_\mathrm{NZ}+H_\mathrm{HF}\, .
\end{equation}
Given a large magnetic field, $H_\mathrm{EZ}(B) \gg H_\mathrm{HF}$, the electron state-dependent nuclear spin splitting difference $\delta_\mathrm{e}=\Delta_{\downarrow_\mathrm{e}}-\Delta_{\uparrow_\mathrm{e}}$ is well approximated by an effective hyperfine constant $\textbf{A} \approx A_\mathrm{eff}$ ($\hat{\textbf{S}}\textbf{A}\hat{\textbf{I}}\approx A_\mathrm{eff}\hat{\textbf{S}} \cdot \hat{\textbf{I}}$) that depends on field direction.

Bergeron \emph{et al.} \cite{Bergeron2020} measured $A_\mathrm{eff}$ for several orientations of the T centre relative to an applied magnetic field, but no experimental determination of the hyperfine tensor has previously been made. In this work, we determine $\textbf{A}$ by resolving the hyperfine structure of an ensemble of T centres in the low and intermediate magnetic field regimes.

\section{Optically-Detected Magnetic Resonance}
 
\begin{figure} [t]
    \centering
    \includegraphics[]{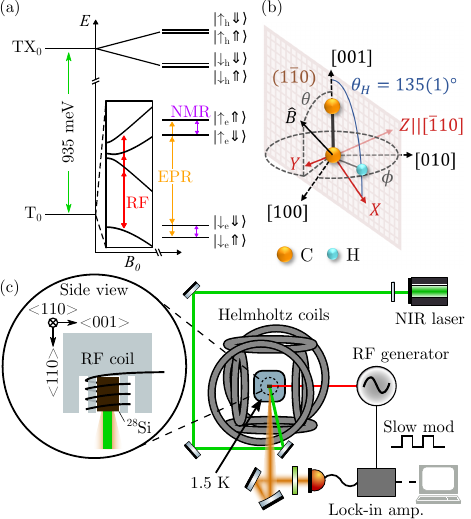}
    \caption{(a) T centre energy level diagram showing the optically detected magnetic resonance (ODMR) schemes. The optical field (green) differentially addresses the ground states. When resonant with a transition, the RF field (red) may transfer population to the `bright' ground state and increase luminescence. At higher fields ($>1$~mT), we simultaneously address an EPR (yellow) and NMR (purple) transition to drive luminescence. Eigenstates are labelled according to the high-field spin composition, though we depict an intermediate field ordering. The sign of $A_\mathrm{eff}$ will determine whether the nuclear spin ordering flips on the $\ket{\uparrow_e}$ or $\ket{\downarrow_e}$ branch compared to high-field. (b) Principal hyperfine axes for orientation z0 \cite{Clear2024optical} shown in red relative to the defect structure and silicon crystal axes (black). The z0 defect plane $(1\Bar{1}0)$ and magnetic field angles $\theta$, $\phi$ are shown. (c) Experiment configuration for ODMR.}
    \label{fig:ODMR exp}
\end{figure}

Optically-detected magnetic resonance (ODMR) allows us to resolve the ground state level structure below the optical resolution \cite{DeAbreu2023Waveguide-integratedCentres}. To characterize the hyperfine structure we perform ODMR with a varying magnetic field along known crystal axes of a T centre ensemble. We resonantly excite an ensemble of T centres at the peak of the zero-field inhomogeneous ZPL using the apparatus shown in \cref{fig:ODMR exp}(c). The steady-state population mixes under resonant excitation, favouring `dark' ground states detuned from the excitation. Because the ground state splittings are within the inhomogeneous distribution for sufficiently small fields ($<1$~mT), the population will polarize into different states for different centres. Resonant radio frequency (RF) driving within the ground state manifold can couple dark and bright spin states, increasing the excited population in steady-state and therefore the sample luminescence.

The sample used in these measurements is the isotopically purified $^{28}$Si crystal with high carbon concentration prepared and measured by Bergeron \emph{et al.} \cite{Bergeron2020}. The T centre sample sits in a liquid He Dewar at 1.5~K. A three turn coil of wire about the sample delivers the RF field (see \cref{fig:ODMR exp}(c)). Three room-temperature Helmholtz pairs are centred on the sample in order to provide a tunable vector magnetic field. A near-infrared laser is aligned at an angle through a window of the cryostat onto the sample, and the resulting photoluminescence is collected out of that same window. Spectral filtering separates laser light from T centre phonon sideband luminescence, which is subsequently measured by a liquid nitrogen cooled Ge diode detector and lock-in amplifier locked to a slow amplitude modulation of the RF signal.
 
At zero-field, we observe that $T_0$ has a close grouping of three higher energy states above a lower-energy fourth state, as shown in \cref{fig:ODMR exp}(a). These differ significantly from the degenerate triplet and singlet states of an isotropic hyperfine coupling. The three highest frequency transitions are 3.482(3)~MHz, 3.713(4)~MHz, and 4.268(3)~MHz. The average observed splitting between these states, 3.821(2)~MHz, is consistent with previous measurements that could not resolve the anisotropy \cite{DeAbreu2023Waveguide-integratedCentres}. Smaller splittings between the closely grouped states can also be seen in the range of $200$--$800$~kHz, as shown in \cref{fig:SpectralCurves}.

With $B > 1$~mT, two RF frequencies are needed to drive luminescence. We apply one RF field in the range of the electron spin splittings to drive electron paramagnetic resonance (EPR) transitions, while the other RF field probes the range of the nuclear magnetic resonance (NMR) transitions.

\section{Hydrogen Hyperfine Tensor}

\begin{figure*}[t]
    \centering
    \includegraphics[]{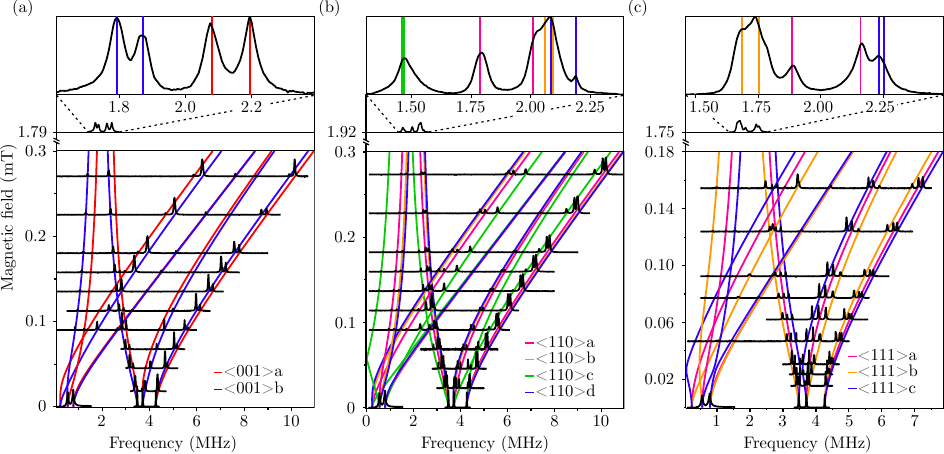}
    \caption{Eigenenergy differences in the T centre ground state as a function of external magnetic field applied along the (a) $<$001$>$, (b) $<$110$>$, and (c) $<$111$>$ crystal axes. The ODMR spectra (black lines) at each field are overlaid by the fit lines (coloured) for each orientation subset, which we label with the field direction and letter a--d. The upper plots show ODMR spectra of the NMR transitions at a  higher field.}
    \label{fig:SpectralCurves}
\end{figure*}

The 12 orientation subsets of the T centre under a general applied magnetic field reduce down to 2, 3, and 4 subsets, respectively, when the field is aligned along the crystal axes $<$001$>$, $<$111$>$, and $<$110$>$ \cite{KAPLYANSKII1967}. The composition of these subsets is given in the Supplemental Material (SM). By performing ODMR with an external field aligned along these symmetry directions, at various field magnitudes, we measure the ground state transitions as a function of $\textbf{B}$. \Cref{fig:SpectralCurves} shows ODMR spectra at fields between 0--2~mT along the three crystal axes. At each field, we fit the observed resonances and extract their peak locations.

We extract the hydrogen hyperfine tensor from the measured ODMR spectra by solving for the eigenenergies of \cref{eq:Tot Ham} and fitting $\textbf{A}$ for all orientation subgroups simultaneously. We use the Euler rotation $R=Z(\gamma)Y(\beta)Z(\alpha)$ (see SM) to generate the set of hyperfine tensors from all orientations in the silicon crystal basis. The principal values of the tensor governing the T centre hydrogen hyperfine coupling are
\begin{equation}
\begin{aligned} 
    &A_X = 4.037(6)\; \text{MHz}, \\
    &A_Y = -4.499(6)\; \text{MHz}, \\
    &A_Z = -2.927(6)\; \text{MHz},
\end{aligned}
\end{equation}
with Euler angles
\begin{equation}
    [\alpha, \beta, \gamma] = [135^\circ, 90^\circ, -45(1)^\circ] \,,
\end{equation}
for the identity orientation z0 as defined by Clear \emph{et al.} \cite{Clear2024optical}. These Euler angles correspond to the hyperfine Z axis pointing normal to the $(1\Bar{1}0)$ defect plane and the X axis rotated an angle $\theta_H=135(1)^{\circ}$ from the $[001]$ crystal axis, as shown in \cref{fig:ODMR exp}(b). We provide the full hyperfine tensor derivation as SM.

Previous work has reported the hydrogen hyperfine tensor based on the electronic structure of the T centre simulated using density functional theory (DFT) \cite{Dhaliah2022}. We determined the value of $\gamma$ from two possible solutions ($-45(1)^\circ$ and $-135(1)^\circ$) by comparison with the defect geometry and these DFT simulations. The positive-value hyperfine axis $X$ points approximately from the bound electron site to the hydrogen nucleus, $\theta_\mathrm{H} = 135^\circ$. We have carried out further analysis to carefully verify the convergence of the hyperfine tensor with respect to the supercell size and K-point sampling, as detailed in the SM. Our results show converged hydrogen principal values of $[A_\mathrm{XX}, A_\mathrm{YY}, A_\mathrm{ZZ}] = [5.347, -4.172, -2.114]$~MHz. We note that though these values qualitatively agree with the experimental measurements in terms of sign and magnitude ordering for the selected rotation case, we observed a difference of up to 35\% compared to the measured values. We hypothesize the discrepancy could come from the absence of vibrational effects including zero-point motion. We discuss these aspects in detail in the SM.

\section{Optically-induced decoherence}

Optically-induced spin decoherence has been widely studied in the diamond NV centre for both the intrinsic nitrogen and host carbon nuclear spins \cite{Jiang2008CoherenceQubit, Wang2015theory, Loenen2025QuantumNetwork, Kalb2018DephasingNetworks}. In this section we provide a similar treatment for the silicon T centre and find magnetic fields that prevent hydrogen memory qubit dephasing.

\begin{figure}
    \centering
    \includegraphics[]{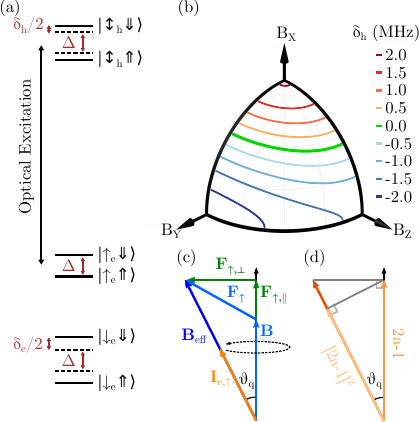}
    \caption{(a) T centre energy levels at high magnetic field. We indicate the arbitrary choice of the optically-coupled hole spin state with $\ket{\updownarrow}$. (b) $\delta_\mathrm{h}$ contours in the high-field limit, shown in the hyperfine basis. The band $\delta_\mathrm{h} = 0$ is asymptotically protected from dephasing during cyclic optical excitation. (c) Evolution of the nuclear magnetic moment $\textbf{I}$ under optical excitation, starting from a ground-state eigenstate $\textbf{I}_{\mathrm{e}, \uparrow}$ aligned along the total effective magnetic field 
    $\textbf{B}_\mathrm{eff}$, which is the sum of the external field and the effective field generated by the the electron. Because TX$_0$ has no hyperfine coupling, the nuclear spin precesses about $\textbf{B}$ during optical excitation. (d) In the long-lifetime limit, the magnetic moment is the average over the precession trajectory. Projection back into the ground state basis returns the original state with probability $[2a-1]^2$, where $a=\braket{\Uparrow_e}{\Uparrow_h}^2$ is the overlap between the ground and excited state nuclear spin up states.}
    \label{fig:TensorOctant}
\end{figure}

Unlike deep defects with transitions between bound states, the T centre's optical transition is between a single-electron ground state and bound exciton excited state. We assume that the hyperfine coupling in TX$_0$ is zero, i.e. $\textbf{A}_\mathrm{h} = 0$, consistent with measurements to date and the large radius of the bound hole ($>1~\mathrm{nm}^3$ \cite{Ivanov2022EffectLocalization}). The nuclear spin splitting and the nuclear spin quantization axis can vary considerably between the two electron spin states of T$_0$ and the hole spin states of TX$_0$, as illustrated in \cref{fig:TensorOctant}(a). We label the difference in nuclear spin splitting between between our reference electron spin state in T$_0$ and the hole spin states of TX$_0$ as $\delta_\mathrm{h}$.

We calculate the evolution of the coupled spins under optical excitation by quantum trajectories. This approach permits analytic solutions, and was tested against an equivalent master equation model. The ground and excited state Hamiltonians are
\begin{eqnarray}
    H_\mathrm{e} & = & -\textbf{B} \left( \gamma_\mathrm{e} \hat{\textbf{S}} + \gamma_\mathrm{n} \hat{\textbf{I}} \right)  + \hat{\textbf{S}} \textbf{A}\hat{\textbf{I}} \,, \\
    H_\mathrm{h} & = &  -\textbf{B} \left( \gamma_\mathrm{h} \hat{\textbf{S}} + \gamma_\mathrm{n} \hat{\textbf{I}} \right) \,,
\end{eqnarray}
where $\gamma_\mathrm{e,h,n}$ are the gyromagnetic ratios of the electron, hole, and the hydrogen nucleus. Since we have assumed $\textbf{A}_\mathrm{h} = 0$, the evolution of the nuclear spin in the excited state is independent of the true electronic structure of TX$_0$ \cite{Clear2024optical}, which we neglect here. For $\gamma_\mathrm{e}=-g_\mathrm{e}\mu_\mathrm{B}/\hbar$ and $\gamma_\mathrm{n}=g_\mathrm{n}\mu_\mathrm{N}/\hbar$ we use $g_\mathrm{e}=2.005$ from Bergeron et al. \cite{Bergeron2020} and the hydrogen nucleus g-factor $g_\mathrm{n}=5.585$. We model optical emission by the partially projective operator $P_\mathrm{he} = \ketbra{h}{e} \otimes \hat{1}_\mathrm{n}$ where $\ket{h}$ and $\ket{e}$ are the spectrally resolved states of the initial TX$_0$ hole state and final T$_0$ electron state respectively and $\hat{1}_\mathrm{n}$ is the identity operator on the nuclear spin subspace. This projection represents optical emission that is too broad to resolve the hyperfine structure,  which can be satisfied by reducing the T centre lifetime, $\tau$, such that $2 \pi \tau \ll 1/\mathrm{max}(A)\approx 35~\text{ns}$.

A trajectory of the system under instantaneous excitation to state $\ket{h}$ at time $t=0$, followed by spontaneous emission and decay to $\ket{e}$ at time $T<t$ is
\begin{equation}
\ket{\psi_\mathrm{traj}(t,T)} = U_\mathrm{e}(t-T) P_\mathrm{eh} U_\mathrm{h}(T) P_\mathrm{he}  \ket{\psi_0} \,,
\end{equation}
with $U_\mathrm{e,h}(t) = e^{-i H_\mathrm{e,h} t/\hbar}$. Given an excited state lifetime of $\tau$, the final state of the system at a time $t \gg \tau$ is a weighted average of trajectories,
\begin{equation}
\rho(t) = \frac{1}{\tau} \int_0^t e^{-T/\tau} \ketbra{\psi_\mathrm{traj}}{\psi_\mathrm{traj}} \, dT \,.
\end{equation}

 This model includes two distinct decoherence mechanisms. First, the projection operators mix the state depending on the degree of entanglement between the electron and nuclear spins. Second, the uncertainty in the time spent in TX$_0$, $\Delta T = \tau$, mixes the final state according to the difference between $U_\mathrm{e}$ and $U_\mathrm{h}$. For lifetimes and magnetic fields relevant to T centre SPIs, $1~\mathrm{\mu s}>\tau>1$~ns and $B>100$~mT, mixing due to $P_\mathrm{he, eh}$ is negligible. Depending on the direction of $\textbf{B}$, the decoherence from $\Delta T$ is predominantly dephasing (Z error on the memory qubit) or relaxation (X error). The memory qubit dephases at a constant rate $\gamma = \delta_\mathrm{h}$, while the memory qubit flip probability $P_\mathrm{flip}(t)$ first increases with time before stabilizing.
 
 In the long lifetime ($\tau\gg1/\abs{\gamma_\mathrm{n}\textbf{B}}$) and secular limits, this model is equivalent to the fluctuator model described by Wang et. al.\cite{Wang2015theory} and illustrated in \cref{fig:TensorOctant}(c) and (d). This model captures the essential mechanism by which optically-induced decoherence can flip the nuclear spin. For example, we consider the evolution of the T$_0$ eigenstate $\ket{\uparrow_\mathrm{e} \Uparrow}$ with nuclear spin $\textbf{I}_0 \approx \textbf{I}_{\mathrm{e},\uparrow} = \tfrac{1}{2} \textbf{B}_\mathrm{eff}/\norm{\textbf{B}_\mathrm{eff}} = \tfrac{1}{2} \vec{b}_\mathrm{eff}$ where the total effective field on the nuclear spin, $\textbf{B}_\mathrm{eff}$, is the sum of the external magnetic field $\textbf{B}$ and the effective field produced by the electron $\textbf{F}_\uparrow$ defined by $\gamma_\mathrm{n} \textbf{F}_\uparrow=\textbf{S}_\uparrow\textbf{A}$ and $\textbf{S}_\uparrow \approx \tfrac{1}{2} \textbf{B}/\norm{\textbf{B}} = \tfrac{1}{2} \vec{b}$ in the secular limit. In the optical excited state, TX$_0$, the spin precesses about $\textbf{B}$ for time $T$ sampled from the excited state lifetime. The mismatch in nuclear quantization axes between the ground and excited states leads to relaxation of the nuclear spin. When the lifetime is long compared to the precession period, the spin averages over the precession and is reduced from $\textbf{I}(t=0) =\tfrac{1}{2} \vec{b}_\mathrm{eff}$ to $\textbf{I}(t\rightarrow\infty) =\tfrac{1}{2} (2a-1) \vec{b}$ where $a=\braket{\Uparrow_\mathrm{e}}{ \Uparrow_\mathrm{h}}^2 = \tfrac{1}{2} (1+\cos{\theta_q})$. Projecting back into the T$_0$ eigenbasis, the long-lifetime limit flip probability is $P_\mathrm{flip}(t\rightarrow \infty) = 4 a(1-a) = \sin^2{(\theta_q)}$, where $\theta_q$ is the angle between $\textbf{B}_\mathrm{eff}$ and $\textbf{B}$.

\begin{figure}
    \centering
    \includegraphics[]{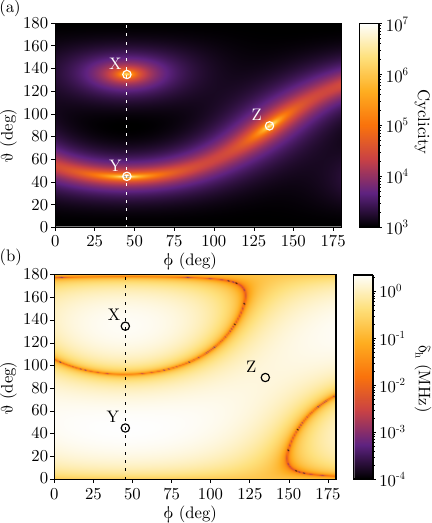}
    \caption{(a) Simulated nuclear cyclicity for T centre orientation z0 with an enhanced lifetime of 10~ns as a function of $\vec{b}$, with $B = 1$~T. State evolution is simulated out to 100~ns. (b) Calculated difference in nuclear spin splitting between the $\ket{\uparrow_e}$ and $\ket{\uparrow_h}$ states ($\delta_\mathrm{h}$). $\theta$ is the polar angle of $\textbf{B}$ from $[001]$ and $\phi$ is the azimuthal angle from $[100]$. The dashed line in both plots represents the defect plane.}
    \label{fig:cyclicity}
\end{figure}

In \cref{fig:cyclicity} we plot the nuclear spin cyclicity $C=1/P_\mathrm{flip}$ and dephasing rate $\delta_\mathrm{h}$ as a function of magnetic field angle for $B=1$~T and with $\tau=10$~ns, which is within reach  of cavity-integrated T centre cavity devices \cite{Afzal2024,Johnston2024Cavity-coupledSilicon, Komza2025MultiplexedArray}. $C$ is maximized along the hyperfine axes where $\textbf{F}_\perp = 0$, while the dephasing rate $\delta_\mathrm{h}$ is zero along a dephasing protection manifold (DPM) about the hyperfine X axis. The coordinates of the DPM in the high-field limit are derived in the SM and plotted in \cref{fig:TensorOctant}(b). Optically-induced decoherence can be reduced below any arbitrary level for some sufficiently large $B$ along the DPM, since $\delta_\mathrm{h} = 0$ and $P^\infty_\mathrm{flip} = \sin^2{\theta_q} \propto 1/B$. The silicon T centre's hydrogen memory qubit can, therefore, be asymptotically protected from optically-induced decoherence.

\section{Discussion}

A SPI possesses a memory qubit DPM like the T centre's only when the effective hyperfine coupling to the memory qubit can be made identical between the ground and optically excited states. If hyperfine coupling is negligible in the excited state, as we assume for the T centre, then the hyperfine coupling tensor must have principal values of opposite sign to satisfy $A_\mathrm{eff} = 0$. This condition requires that the ground state isotropic contact hyperfine is smaller than the anisotropic, zero-trace dipole hyperfine coupling. For popular SPIs, such as the NV, SiV and SnV diamond colour centres \cite{Harris2023HyperfineSpectroscopy}, this condition is met only by remote nuclear spins \cite{Dreau2014ProbingDynamics}. In this regard the T centre's ground state H hyperfine coupling tensor is more similar to the ground state of some rare earth ions, whose intrinsic nuclear spins can hyperfine couple to the 4f electron orbitals with dipolar character \cite{Welinsky2016HighResolution}. Unlike these ions, the T centre's electron Lande g-factor $g_\mathrm{e}$ is isotropic (to within 1\% \cite{Clear2024optical}).
 
Capitalizing on the DPM to reduce decoherence requires a sufficiently high magnetic field. Of course, in practical quantum information devices the maximum field is constrained and there exists some finite decoherence in the form of relaxation even along the DPM. Moreover, $\delta_\mathrm{e}\rightarrow0$ along the DPM and it may be desirable to further constrain $\vec{B}$ so that the electron transitions are resolved. $\delta_\mathrm{e}$ reaches a maximum of $106~\text{kHz}$ along the $\delta_\mathrm{h} =0$ manifold for $B = 1$~T, as shown in \cref{fig:cyclicity}(b).

Here we consider strategies to further limit decoherence at finite fields, and how they inform the choice of magnetic field direction. The most critical approach is to reduce the optical excited state lifetime by Purcell enhancement e.g. with an optical nanocavity \cite{Johnston2024Cavity-coupledSilicon, Islam2023cavityenhanced, Afzal2024, Komza2025MultiplexedArray}. The shorter the lifetime, the less optically induced decoherence per optical cycle. For simplicity, we first consider $\textbf{B} \parallel Z$ (the T centre symmetry axis). In this configuration optically induced relaxation is negligible ($C\rightarrow \infty$ in \cref{fig:cyclicity}). At $\tau=10$~ns ($1$~ns), $\delta_\mathrm{h} = 2 \pi \times 300$~kHz ($3$~MHz) equates to an infidelity of $10^{-4}$ per optical cycle. This demonstrates that per-cycle optically induced decoherence can be reduced far below the thresholds for quantum error correction codes with feasible T centre lifetimes, even outside of the DPM.

Performance can be further improved by photon detection and feedback. Since the decoherence is primarily due to uncertainty in the time spent in the optically excited state TX$_0$, a unitary $U^\mathrm{det}_\mathrm{corr} = U_\mathrm{e}^\dag(t-T) U_\mathrm{h}^\dag(T)$ can be applied to the memory qubit to correct for its differential evolution if the emission time is known. If, for example, half of the emitted photons are collected during readout and the corresponding $U_\mathrm{corr}$ applied, then the average decoherence is halved. Integrated T centres to date are lossy, and so the proportion of emission events that can be precisely corrected is very low.

However, an average unitary correction $U^\mathrm{avg}_\mathrm{corr}$ can be applied to improve the memory fidelity even when the photon is undetected, and the emission time $T$ is unknown, so long as the initial and final electron states are known. $U^\mathrm{avg}_\mathrm{corr}$ corrects for the average evolution of the mixed system according to
\begin{align}\label{eqn:unitary_corr}
U^\mathrm{avg}\left(t, \tau\right)& = \frac{1}{\tau} \int e^{(-T/\tau)} U_\mathrm{e}(t-T) U_\mathrm{h}(T) \,dT  \,, \\
U^\mathrm{avg}_\mathrm{corr}\left(t, \tau\right)& = \frac{ U_\mathrm{e}(t) \left(U^\mathrm{avg}\left(t, \tau\right)\right)^\dag}{\mathrm{Norm}\left[U^\mathrm{avg}\left(t, \tau\right) \right] } \,.
\end{align}
This corrects for a portion of both the dephasing and relaxation without measurement dependence. The resulting fidelity is limited only by the purity of $\rho(t)$. 

For $\textbf{B} \parallel Z$, $U^\mathrm{avg}_\mathrm{corr} = \mathrm{Diag}\left[-\delta_\mathrm{h} \tau, +\delta_\mathrm{h} \tau\right]$, which is a Z rotation (or frame update), and halves the short-lifetime infidelity from $I= 1-F =\tfrac{1}{2} \left(\tau \delta_\mathrm{h}\right)^2$ to $I_\mathrm{corr} = \tfrac{1}{4} \left(\tau \delta_\mathrm{h}\right)^2$. In the long-lifetime limit, the memory qubit completely dephases and cannot be recovered. For $\vec{B}/\abs{B} \not\in  X,Y,Z$ (or any hyperfine axis) $U^\mathrm{avg}_\mathrm{corr}$ consists of both Z and X rotations. 

Along the DPM in \cref{fig:cyclicity}, average unitary correction reduces $P_\mathrm{flip}(t=\infty) = \sin^2{\left(\theta_q \right)}$ to $P_\mathrm{flip, corr}(t=\infty) = 2 \sin^2{\left( \theta_q/2 \right)}$, in the high-field limit. We present the calculated nuclear spin cyclicity at finite fields, including optimal average unitary correction, as a function of $\vec{B}$ in the SM. This is the toolkit required to choose an optically-enhanced lifetime, magnetic field magnitude, and magnetic field direction to achieve the desired memory qubit performance with silicon T centre SPIs.

This simple model can be extended in a number of ways. For example, it can include branching to the unwanted electron spin state with a projection operator weighted by the electron branching ratio \cite{Clear2024optical}. Here we assume electron branching is made negligible by selective Purcell enhancement. The projection operators can also be made partially resolving to capture decoherence induced by optically resolved hyperfine states.

\section{Conclusion}

We have made precise measurements of the silicon T centre's ground-state structure in the low to intermediate magnetic field regime and determined the anisotropic hyperfine coupling between the T centre's bound electron and hydrogen nucleus for the first time. We have further demonstrated the existence of a dephasing protection manifold which permits a choice of external magnetic field such that optically-induced nuclear spin decoherence vanishes asymptotically with increasing $B$---suitable for brokered entanglement between T centres. The techniques we propose may enable high-fidelity entanglement between cavity-coupled T centres in silicon photonic circuits, and over fibre networks for networked quantum computing and secure quantum communication \cite{Simmons2024}.

The same experimental techniques can be used to characterize the hyperfine coupling between the bound electron and carbon nuclei in the T centre's $^{13}$C isotopic variants. These T centres with four-qubit local spin registers will allow more sophisticated local quantum logic such as entanglement distillation \cite{Nickerson2014FreelyLinks, Waldherr2014}, and potentially extend coherence times further through decoherence-free subspaces and clock transitions \cite{Reiserer2016Robust}.
\newline
\newline
See Supplemental Material for supporting content.
\newline
\begin{acknowledgments}

This work was supported by the Natural Sciences and Engineering Research Council of Canada (NSERC), the New Frontiers in Research Fund (NFRF), the Canada Research Chairs program (CRC), the Canada Foundation for Innovation (CFI), the B.C. Knowledge Development Fund (BCKDF), and the Canadian Institute for Advanced Research (CIFAR) Quantum Information Science program. 

The work on first principles computations was supported by the U.S. Department of Energy, Office of Science, Basic Energy Sciences in Quantum Information Science under Award Number DE-SC0022289. This research used resources of the National Energy Research Scientific Computing Center, a DOE Office of Science User Facility supported by the Office of Science of the U.S. Department of Energy under Contract No. DE-AC02-05CH11231 using NERSC award BES-ERCAP0020966.

The $^{28}$Si samples used in this study were prepared from the Avo28 crystal produced by the International Avogadro Coordination (IAC) Project (2004–2011) in cooperation among the BIPM, the INRIM (Italy), the IRMM (EU), the NMIA (Australia), the NMIJ (Japan), the
NPL (UK), and the PTB (Germany).

\end{acknowledgments}

\end{document}